\documentclass[aps,prc,superscriptaddress,showpacs,nofootinbib,floatfix,twocolumn]{revtex4}

\usepackage{doi,graphicx,amssymb,amsmath,dcolumn}

\newcommand{\unit}[1]{\ensuremath{\, \mathrm{#1}}}
\newcommand{\etal}{\textit{et al.}}
\newcommand{\ie}{\textit{i.e.}}
\newcolumntype{d}[1]{D{.}{.}{#1} }
\begin{document}

\title{Parking-garage structures in astrophysics and biophysics}
\author{C. J. Horowitz}\email{horowit@indiana.edu}
\author{D. K. Berry}\email{dkberry@iu.edu}
\author{M. E. Caplan}\email{mecaplan@indiana.edu}
\affiliation{Center for Exploration of Energy and Matter and Department of Physics, Indiana University, Bloomington, IN 47405, USA}
\author{Greg Huber}\email{huber@kitp.ucsb.edu}
\affiliation{Kavli Institute for Theoretical Physics, Kohn Hall,
University of California, Santa Barbara, California 93106-4030, USA}
\author{A. S. Schneider}\email{andschn@indiana.edu}
\affiliation{TAPIR, California Institute of Technology, Pasadena, CA 91125, USA}

\date{\today}
\begin{abstract}
A striking shape was recently observed for the cellular organelle endoplasmic reticulum consisting of stacked sheets connected by helical ramps \cite{Cell154:285}.  This shape is interesting both for its biological function, to synthesize proteins using an increased surface area for ribosome factories, and its geometric properties that may be insensitive to details of the microscopic interactions.  In the present work, we find very similar shapes in our molecular dynamics simulations of the nuclear pasta phases of dense nuclear matter that are expected deep in the crust of neutron stars.   There are dramatic differences between nuclear pasta and terrestrial cell biology.  Nuclear pasta is 14 orders of magnitude denser than the aqueous environs of the cell nucleus and involves strong interactions between protons and neutrons, while cellular scale biology is dominated by the entropy of water and complex assemblies of biomolecules.  Nonetheless the very similar geometry suggests both systems may have similar coarse-grained dynamics and that the shapes are indeed determined by geometrical considerations, independent of microscopic details.   Many of our simulations self-assemble into flat sheets connected by helical ramps.   These ramps may impact the thermal and electrical conductivities, viscosity, shear modulus, and breaking strain of neutron star crust.   The interaction we use, with Coulomb frustration, may provide a simple model system that reproduces many biologically important shapes.

\end{abstract}


\pacs{26.60.-c,02.70.Ns, 87.10.Tf, 87.15.ap, 87.16.D-,87.16.Tb}

\maketitle


Membrane-bound cellular organelles have characteristic shapes, with the endoplasmic reticulum (ER) being particularly striking.  The ER is an extensive organelle displaying three distinct, yet connected, morphologies: tubes, sheets and the spherical envelope around the cell nucleus.  Recent advances in serial sectioning and electron microscopy have revealed the stacked ER sheets to be connected by helical structures \cite{Cell154:285}.  Just as spiral ramps connect the levels of a multilevel parking garage, these so-called ``Terasaki ramps" allow sheets to connect yet remain parallel over scales large relative to the membrane thickness. This parking-garage shape is interesting for both its biological function, to synthesize proteins using an increased surface area for ribosome factories, and its mathematical property as the minimizer of a geometrical Hamiltonian largely insensitive to details of the microscopic interactions \cite{GuvenPRL}.  In these models of membrane mechanics, the Terasaki ramps are stable, topological structures, akin to screw dislocations, that affect the entire morphology of an organelle critical to the metabolism of eukaryotic cells.

Independent of considerations from terrestrial biology, in this Letter we find very similar shapes arising in our molecular dynamics (MD) simulations of dense nuclear phases.  These phases are expected in neutron-star crusts and in core-collapse supernovae \cite{PRL114:031102}.  If matter is compressed to near nuclear densities, competition between short-range nuclear attraction and long-range Coulomb repulsion can lead to a variety of complex shapes, so-called ``nuclear pasta" \cite{Pasta,Pasta2}.  Note that Groenewold \etal ~\cite{gron2004} discuss colloidal cluster phases arising from short range attraction and coulomb repulsion.  For a particular density and proton fraction, we also find flat sheets connected by spiral ramps.  
In a neutron star, electrons scattering from the spiral ramps reduce both the thermal and electrical conductivity.  This may impact X-ray observations of crust cooling in transiently accreting stars \cite{PRL114:031102}, and may also lead to the decay of neutron-star magnetic fields after about a million years \cite{Pons}.    

There are, to be sure, dramatic differences between nuclear pasta and terrestrial cell biology.  Nuclear pasta has a density near $10^{14}$ g/cm$^3$, fully fourteen orders of magnitude more dense than the aqueous environs of the cell nucleus.  Furthermore, nuclear pasta involves strong interactions between neutrons and protons in addition to electromagnetic interactions, while cellular-scale biology is highly-screened, highly-overdamped and dominated by the entropy of water and complex assemblies of biomolecules.  Nonetheless the strikingly similar geometry suggests both systems may have similar coarse-grained dynamics.

In nuclear physics, the semi empirical mass formula predicts the binding energy $BE$ of a nucleus with $A$ nucleons (neutrons plus protons) and $Z$ protons \cite{Wei1935, Mol1988},
\begin{equation}
BE=a_v A - a_s A^{2/3} - a_c Z^2/A^{1/3}...
\label{eq.semf}
\end{equation}
Here $a_v$, $a_s$, $a_c$ are constants describing volume, surface, and Coulomb energies.  In addition there are other contributions from the symmetry energy and pairing that will not be important here.  Competition between surface and Coulomb energy contributions can lead to complex shapes.  We perform simulations at a density of $n=0.05$ fm$^{-3}$.  This corresponds to a packing fraction of 5/16 of nuclear saturation density, $n_0=0.16$ nucleons per fm$^3$.  Here the system forms flat sheets (lasagne) that are considerably thicker than the size of a single nucleon.  This thickness is determined from a balance of surface and Coulomb energies.  
Note that there are a variety of different shapes with almost the same energy.  For example, if the density is decreased somewhat, the system forms rods (spaghetti) instead of flat sheets and at still lower densities the system forms spheres representing isolated nuclei.

Fluctuations about these simple shapes could have low excitation energies that may depend on sub-domaint terms in Eq. \ref{eq.semf}.  Reinhard \etal \ \cite{PRC73:014309}, see also \cite{ZPhysA346:87}, use a leptodermous (thin-skinned) expansion and density functionals in order to calculate a curvature energy term for Eq. \ref{eq.semf} that goes like $a_{curve}A^{1/3}$.   The impact of such an $A^{1/3}$ term on nuclear pasta shapes was considered in ref. \cite{curvedpasta}.  The curvature energy could be important for nuclear fission where a nucleus dramatically changes shape.  However, Reinhard \etal\ find they need to calculate the energy of very large nuclei with $A$ in the thousands in order to extract $a_{curve}$ theoretically.  Therefore, this coefficient  may not be reliably determined from measured nuclear binding energies that are known only for a limited range of $A$.

Instead, we discuss a very different approach that has been employed previously in biophysics, but not (to our knowledege) in nuclear physics.   We consider a Helfrich-Canham Hamiltonian $H_0$  \cite{Helfrich, Canham} that involves a quadratic form in the surface principal curvatures $C_1$ and $C_2$ including both the mean curvature $(C_1+C_2)/2$ and Gaussian curvature $C_1C_2$,
\begin{equation}
H_0=\frac{1}{2}B\int dS (C_1+C_2)^2 + \bar B \int dS C_1 C_2\,.
\label{eq.helfrich}
\end{equation} 
Here $\int dS$ is an integral over the surface area and $B$ and $\bar B$ are constants representing effective rigidity moduli.  This energy functional was applied in refs. \cite{GuvenPRL,Cell154:285} to biological systems. Our justification for also discussing nuclear pasta shapes with the help of Eq. \ref{eq.helfrich} is determined only after the fact.  This equation predicts spiral ramp shapes, and their arrangements, very similar to what we find in our MD simulations, see below.  

One relatively low energy solution for Eq. \ref{eq.helfrich} is lasagne with flat surfaces where $C_1=C_2=0$. This gives $H_0=0.$  Another solution involves spiral ramps with $C_1=-C_2$.  Here the mean curvature is still zero, but the Gaussian curvature is negative so that $H_0<0$.  Thus the Gaussian curvature term may stabilize spiral ramp configurations in both biological membranes and nuclear pasta.  This term may also stabilize configurations that have additional holes such as the ``nuclear waffle'' shapes found in ref. \cite{Waffles}.  These shapes consist of flat sheets with a two dimensional array of holes.  Similarly, the ER displays a waffle-type pattern which has been referred to in the biological literature as ``fenestrations" \cite{Puhka}.  Alternatively there may be torus or donut-shaped superheavy nuclei \cite{nuclear_donuts} where the donut shape both reduces the large Coulomb energy and is further stabilized by the Gaussian-curvature term. 
  
In this paper we study the self-assembly of these spiral ramp configurations with molecular dynamics simulations of a simple (semi)classical model of nuclear matter.  Our simulations are for nuclear pasta but they may also have implications for phospholipid bilayer membranes.


Our MD formalism is the same as that used by Horowitz \etal \  in previous works 
\cite{PRL114:031102,PhysRevC.69.045804,PhysRevC.70.065806,PhysRevC.72.035801,PhysRevC.78.035806,PhysRevC.88.065807} and is briefly reviewed here.   It is very similar to a model used by others \cite{PhysRevC.86.055805,PhysRevC.89.055801}.  Our simulation volume is a cubic box with periodic boundary conditions which contains point-like protons and neutrons with mass $M=939\unit{MeV}$. Electrons are assumed to form a degenerate relativistic Fermi gas and are not explicitly included in the simulations. Protons and neutrons interact via the two-body potentials:
\begin{subequations}
\begin{align}
 V_{np}(r)&=a\, {\rm e}^{-r^2/\Lambda}+b\, {\rm e}^{-r^2/2\Lambda}\\
 V_{nn}(r)&=a\, {\rm e}^{-r^2/\Lambda}+ c\, {\rm e}^{-r^2/2\Lambda}\\
 V_{pp}(r)&=a\, {\rm e}^{-r^2/\Lambda}+ c\, {\rm e}^{-r^2/2\Lambda}+\frac{\alpha}{r}{\rm e}^{-r/\lambda}.
\end{align}
\label{eq.pot}
\end{subequations}
The $n$ and $p$ indices indicate whether the potential describes a neutron-proton, a neutron-neutron, or a proton-proton interaction. Meanwhile, $r$ is the separation between each pair of interacting nucleons and quantities $a=110$ MeV, $b=-50$ MeV, $c=-2$ MeV, and $\Lambda=1.25$ fm$^2$ are parameters of the model. Their values were chosen in ref. \cite{PhysRevC.69.045804} to approximately reproduce some bulk properties of pure neutron matter and symmetric nuclear matter, as well as the binding energies of selected nuclei.  The screening length $\lambda$ is chosen to be 10 fm.  Equation \ref{eq.pot} describes an intermediate range attraction between $p$ and $n$ which binds nuclei and then a short range repulsion that causes nuclear matter to saturate at a density $n_0=0.16$ fm$^{-3}$.  Finally there is a long range Coulomb repulsion between protons.  All of the simulations in this paper are at a density of $n=0.05$ fm$^{-3}$, a composition of 40\% protons, 60\% neutrons, a fixed temperature $kT=1$ MeV, and use a MD time step of 2 fm/c. 


\begin{figure}[h]
\centering
\includegraphics[width=0.51\textwidth]{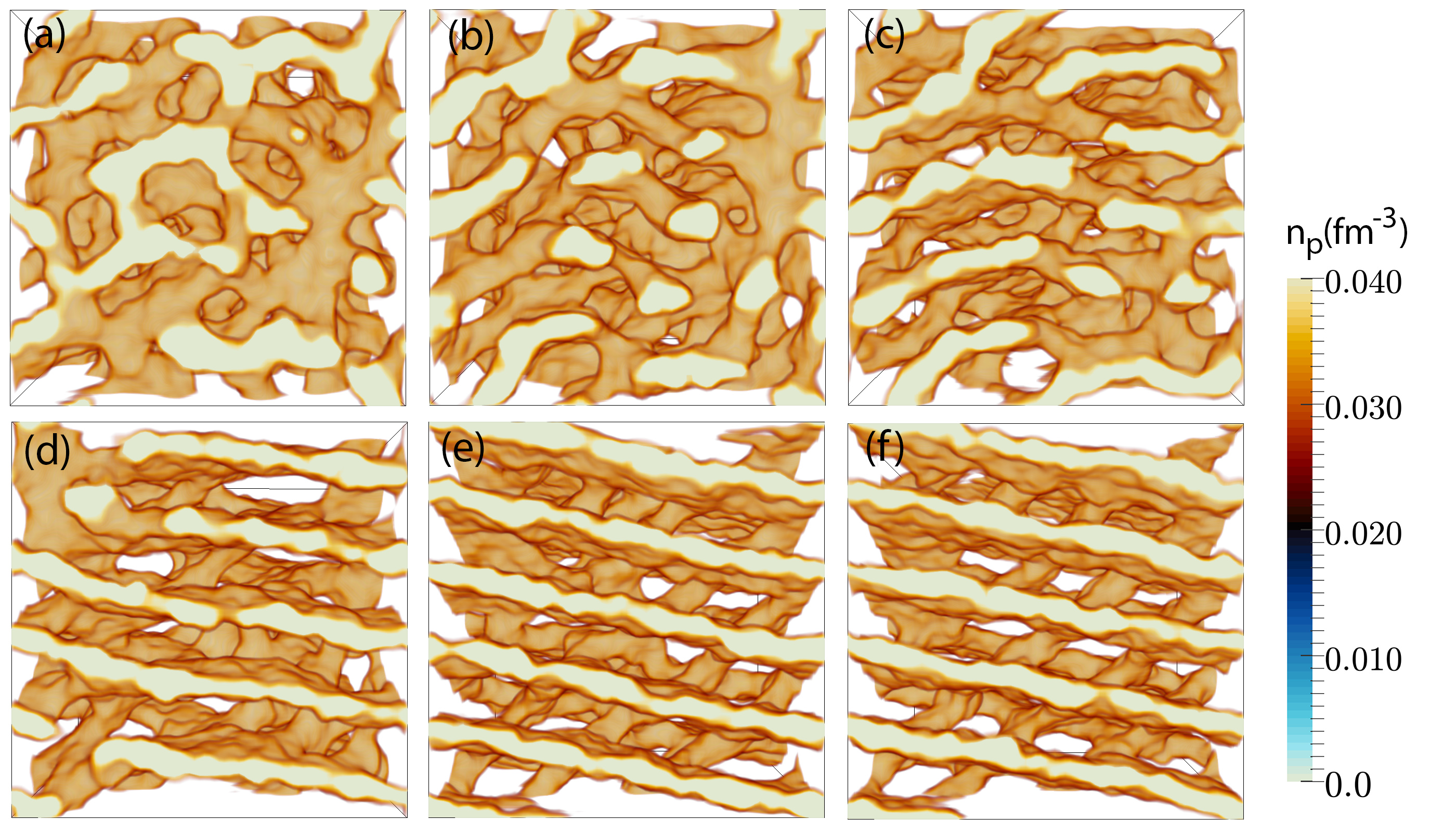}
\caption{(Color online) Self-assembly of a parking garage structure in a MD simulation with 40,000 nucleons that started from uniform random initial positions.  Shown in panels (a) through (f) are the configuration after simulation times of 40,000, 400,000, 800,000, 1,200,000, 1,600,000 and 2,000,000 fm/c respectively.  The color map at right is discussed in the text.  Figure produced with Paraview \cite{Paraview}.}
\label{fig:self_assembly}
\end{figure}

\begin{figure}[h]
\centering
\includegraphics[width=0.5\textwidth]{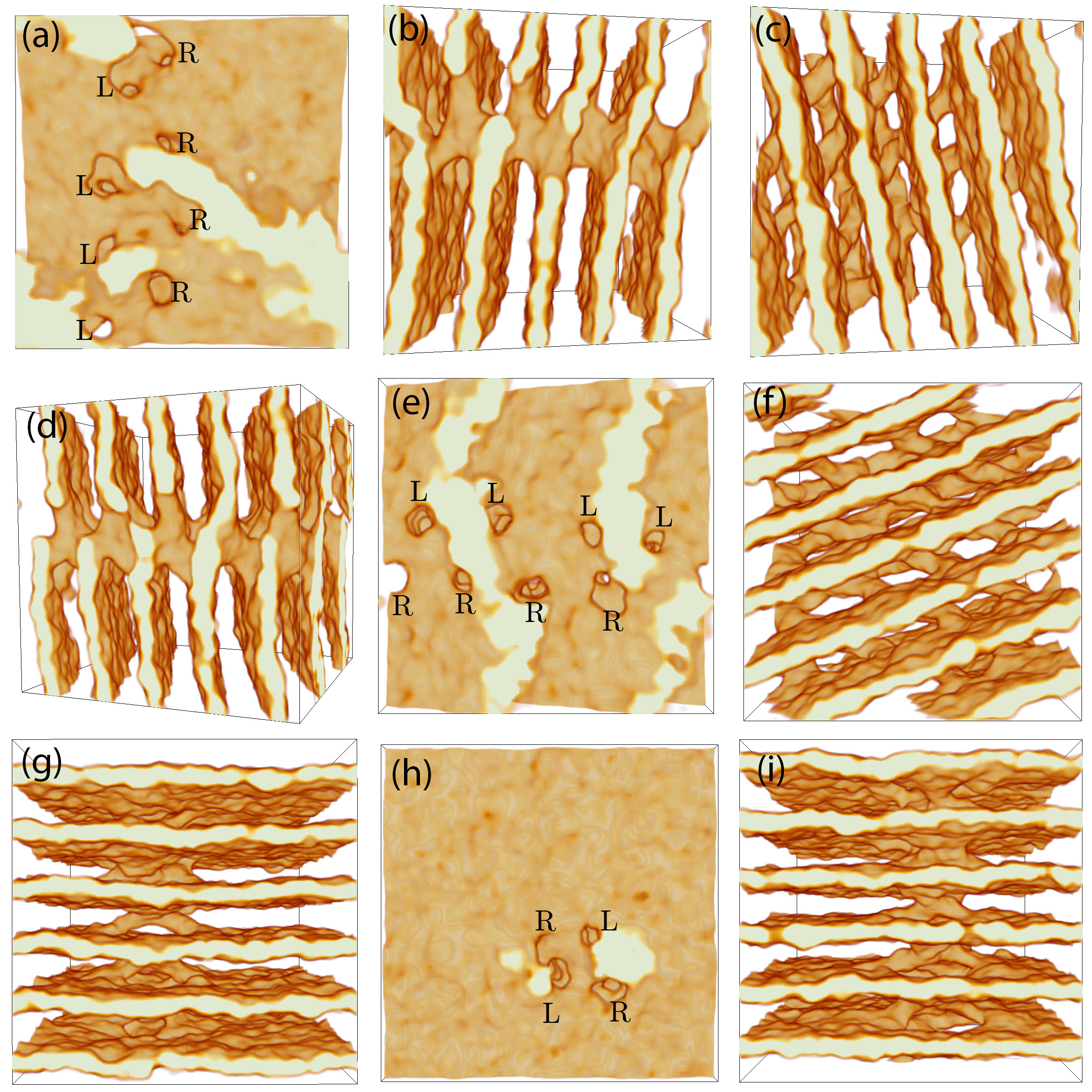}
\caption{(Color online)  Three axis views, X (left), Y(center), and Z (right), of the final configuration of three MD simulations.  Panels (a)-(c) at top are for the 40,000 nucleon simulation shown in Fig. 1.  Panels (d)-(f) at center are for a simulation with 50,000 nucleons and panels (g)-(i) at bottom are for a simulation with 75,000 nucleons.  The handedness of the helical holes in panels (a), (e), and (h) are indicated with L for left-handed and R for right-handed.}
\label{fig:XYZ}
\end{figure}

To study self-assembly of these ramps, we start from initial conditions where the particle positions are uniformly distributed in the simulation volume with a Boltzmann velocity distribution.    Figure \ref{fig:self_assembly} shows the configuration of a 40,000 particle simulation at times from 40,000 to 2,000,000 fm/c.  The proton density $n_p$ is shown where each proton is assumed to have a Gaussian shape of 1 fm$^3$ volume \cite{Waffles} and the opacity of the colorscale is 0 for $n_p = 0.00$ to 0.02 fm$^{-3}$ and then increases linearly to 1 at $n_p=0.04$ fm$^{-3}$.   A light cream color corresponds to high density sheets, while lower density surfaces are shown in brown.  This system undergoes the following self-assembly steps:  (1) the low density system collapses locally to form higher density filaments that meet in junctions, see Fig. \ref{fig:self_assembly} (a).  This includes the formation of a number of topological holes. (2) Next, the filaments start to grow to form curved sheets, Fig. \ref{fig:self_assembly} (b-c), (3) these sheets then start to straighten out over longer length scales, Fig. \ref{fig:self_assembly} (d). (4) Boundaries between ``domains'' of sheets with different orientation form four left-handed and four right handed helical ramps, see Table \ref{tab:one}.   The sheets straighten out over the full simulation volume, see Fig. \ref{fig:self_assembly} (e-f), and the  ramps move together to form the dipole pattern shown in Fig. \ref{fig:XYZ} (a).  This pattern has four left-handed ramps to the left and four right-handed ramps to the right.  For this dipole pattern, the ramps are seen in Fig. \ref{fig:XYZ} (c) to make about a 45 degree angle with the flat sheets.  We find this final configuration to be stable for times of at least 10,000,000 fm/c.  

To study the dependence on boundary conditions, we perform simulations with different numbers of particles and correspondingly different sized simulation volumes, see Table \ref{tab:one}.  The smallest simulation, with only 20,000 particles, forms uniform flat sheets without any spiral ramps (not shown).  A 50,000 particle simulation forms four left handed and four right handed ramps as shown in Fig. \ref{fig:XYZ} (e).  This is also a dipole pattern with the left-handed ramps to the left and the right-handed ramps to the right.  As a result the ramps make a 45 degree angle with the sheets, see Fig. \ref{fig:XYZ} (f).  Finally a 75,000 particle simulation forms four ramps in a quadrupole pattern where one left-handed and one right-handed ramp are to the left as shown in Fig. \ref{fig:XYZ} (h).  For a quadruple pattern the ramps are observed to make a 90 degree angle with the sheets, see Fig. \ref{fig:XYZ} (i) and Table \ref{tab:one}.

The arrangement of the helical ramps shown in Fig. \ref{fig:XYZ} panels (a), (e) and (h) agrees very well with theoretical predictions in refs. \cite{Cell154:285, GuvenPRL} that are based on Eq. \ref{eq.helfrich}.  Guven \etal\ argue that tension in the sheets leads to an effective long range attraction between two ramps of opposite chirality that draws them together until a short range repulsive bending force stabilizes the pair of ramps at a characteristic distance.  Some features of our simulations that are in agreement with Guven \etal\ are: (1) we find an equal number of left-handed (L) and right-handed (R) ramps, (2) the ramps are all relatively close together with a similar characteristic spacing, and (3) ramps in a quadrupole pattern make a 90 degree angle with respect to the sheets while ramps in a dipole pattern make an approximately 45 degree angle with the sheets.  These common features suggest, after the fact, that Eq. \ref{eq.helfrich} may also apply to our nuclear pasta model system. 

\begin{table}[h]
\caption{\label{Tab:parameters} Number of ramps $N_r$ in MD simulations with different numbers of particles $N$. 
}
\begin{ruledtabular}
\begin{tabular}{*{4}{c}}
$N$ &$N_r$ & Pattern of ramps & Angle of ramps (deg) \\
  20,000    &  0    &       &          \\
  40,000   &  8     & Dipole & 45 \\
  50,000   &  8     & Dipole  & 45 \\
  75,000   &  4     & Quadrupole & 90 \\ 
\end{tabular}
\end{ruledtabular}
\label{tab:one}
\end{table}

Most of our simulations form flat sheets connected by helical ramps.  However we are able to obtain only flat sheets, without any ramps, if we add a small one-body potential to the system for early times that biases the formation of only flat sheets, or start a simulation in a smaller box at high densities where the system is nearly uniform and then very slowly expand the box during the simulation until the system reaches the same final density of $n=0.05$ fm$^{-3}$.

\begin{figure}[h]
\centering
\includegraphics[width=0.5\textwidth]{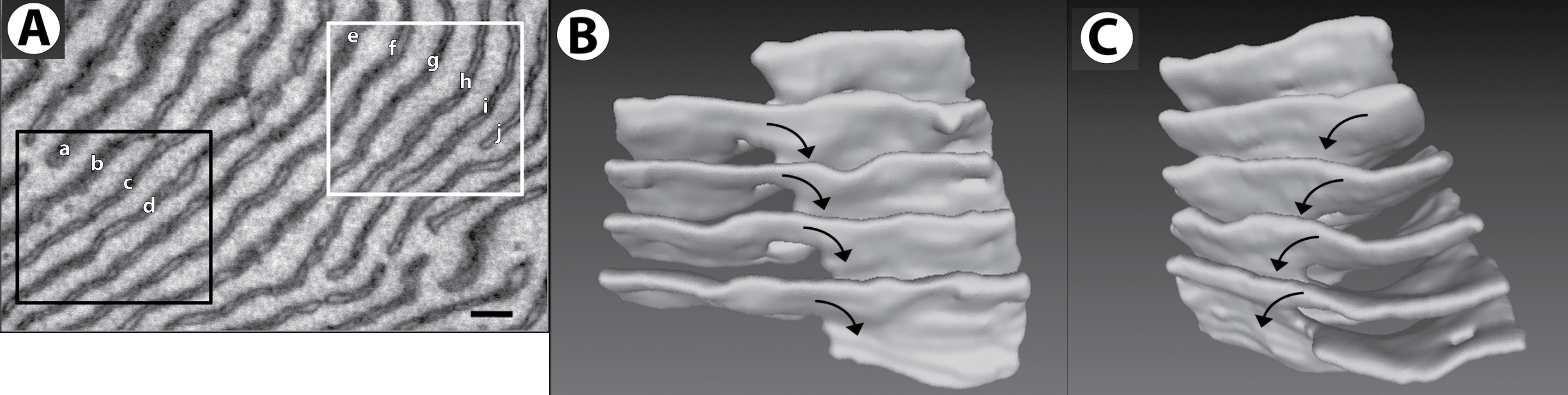}
\caption{(A) Scanning electron micrograph of a thin slice of sheet-like ER from a mouse salivary gland.  Serial sectioning allows three-dimensional reconstruction from large numbers of cross-sections (scale bar = 200 nm).   (B) 3D reconstruction of a left-handed Terasaki ramp that appears in the black-outlined region in (A).  (C) 3D reconstruction of a right-handed Terasaki ramp that appears in the white-outlined region in (A).  (Figure adapted from Terasaki et al. \cite{Cell154:285})}
\label{fig:images}
\end{figure}

We compare our results to biological observations.  Reconstructing the three-dimensional geometry of ER sheets from two-dimensional serial sections reveal that the continuity of parallel sheets comes about through the helical winding of the ``exposed" sheet edge through space, \ie \ the Terasaki ramp, see Fig. \ref{fig:images} .  The core of the ramp (cytosolic side of the membrane) has a highly negative Gaussian curvature, but potentially small mean curvature, given the opposite signs of the two principal radii of curvature.  No considerations seem to set a preferred handedness to the ramps, and the observations at hand, though statistically small, are consistent with right- and left-handed ramps in equal numbers throughout the organelle.  If the sheet edges are treated as effectively one-dimensional defects (thought to be stabilized by membrane proteins on the cytosolic side), then it is sufficient to treat the rest of the membrane by the Helfrich-Canham Hamiltonian.
A class of solutions minimizing that functional are minimal surfaces, which have zero mean curvature and locally minimize area.  One consequence of the analysis in \cite{GuvenPRL} is that whereas a single Terasaki ramp has a logarithmically diverging energy, a left-right dipole pair has a finite energy, and a double pair of left-right-left-right ramps (a quadrupole) minimizes it further.  The existence of ramp dipoles is natural in the biological context, though convincing evidence for tightly correlated left-right pairs, or for pairs of pairs, is currently lacking.  As previously discussed, we find both left-right pairs and pairs of pairs in our MD simulations of nuclear pasta.


In conclusion, we have performed molecular dynamics simulations using a simple classical model of nuclear pasta.  Many of our systems spontaneously self-assemble to form flat sheets connected by helical ramps.  This geometry is very similar to that observed in the three-dimensional structure of the sheet-like endoplasmic reticulum.  Seeing the same helical shapes in the extraordinarily different systems of nuclear pasta in neutron stars, and in the membranes of eukaryotic cells strongly suggests that these shapes follow from common geometric considerations and may be independent of details of the microscopic interactions.

The curvature energy could help stabilize these helical ramps, suggesting that they may be common in nuclear pasta.  If true, electron scattering from the ramps could reduce the electrical and thermal conductivity of neutron-star crust.  This may impact crust cooling \cite{PRL114:031102} and possibly lead to the decay of magnetic fields \cite{Pons}.   Alternatively the helical ramps, that connect flat sheets, may allow paired protons to percolate throughout the system and make it superconducting.  In addition, the ramps may restrict how the sheets can move past each other.  This could lead to complex viscoelastic behavior for nuclear pasta that could dampen r-mode oscillations in rapidly rotating neutron stars, see for example \cite{rmode}.  Furthermore, the helical ramps likely will increase the shear modulus of neutron-star crust and the frequencies of crust oscillation modes.  These modes may have been observed as quasi periodic oscillations during magnetar giant flares \cite{magnetar}.  Finally, the ramps likely increase the breaking strain, or mechanical strength, of neutron star crust.  As a result the crust can support larger mountains that, on a rapidly rotating neutron star, efficiently radiate gravitational waves \cite{breakingstrain}.


One can study soft condensed-matter analogs of nuclear pasta in the laboratory even if direct experiments on nuclear pasta are not feasible.  This could provide unique insights.  For example, it may be very difficult to predict the actual density of ramps in nuclear pasta using only first principle simulations.  Instead one can observe the actual density and pattern of Terasaki ramps in analog laboratory systems.

Our simple interaction has Coulomb frustration with short-ranged attraction and long-ranged repulsion.  This seems to provide a simple model system that reproduces many biologically important shapes.   One reason is the fluid nature of bilayer membranes: interactions between phospholipid molecules are determined to a large extent by the repulsive term.

Computational advances have made and will make very large-scale MD simulations for such systems ``easy'' and these simulations will likely exhibit very rich varieties of shapes and phases.  Uncovering similarities between disparate physical systems allows connections to be made at the deeper level of symmetry, excitations and the geometry of topological defects.

\begin{acknowledgments}

We thank Gerardo Ortiz and Sima Setayeshgar for helpful comments.  Part of this work was completed at the Aspen Center for Physics, which is supported by National Science Foundation grant PHY-1066293. This research was supported in part by DOE grants DE-FG02-87ER40365 (Indiana University) and DE-SC0008808 (NUCLEI SciDAC Collaboration), and NSF grant PHY11-25915 (KITP, UCSB).

\end{acknowledgments}


\begin{thebibliography}{99}
\bibitem{Cell154:285} M. Terasaki {\it et al.}, Cell {\bf 154}, 285 (2013).

\bibitem{GuvenPRL} J. Guven, G. Huber, and D. M. Valencia, Phys. Rev. Lett. {\bf 113}, 188101 (2014).

\bibitem{PRL114:031102}C. J. Horowitz, D. K. Berry, C. M. Briggs, M. E. Caplan, A. Cumming, and A. S. Schneider, Phys. Rev. Lett. {\bf 114}, 031102 (2015). 

\bibitem{Pasta}D. G. Ravenhall, C. J. Pethick, and J. R. Wilson, Phys. Rev. Lett. {\bf 50}, 2066 (1983).

\bibitem{Pasta2} M. Hashimoto, H. Seki, and M. Yamada, Progress of Theoretical  Physics {\bf 71},  320  (1984).
 
 \bibitem{gron2004} J. Groenewold, W. K. Kegel, J. Phys: Condens. Matter {\bf 16}, S4877 (2004).

\bibitem{Pons}J. A. Pons, D. Vigano, and N. Rea, Nat. Phys. {\bf 9}, 431 (2013).

\bibitem{Wei1935} C. F. von Weizacker, Zeitschrift fur Physik {\bf 96}, 431 (1935).

\bibitem{Mol1988} P. Moller, W. D. Myers, W. J. Swiatecki, and J. Treiner, Atomic data and nuclear data tables, {\bf 39}, 225 (1988).

\bibitem{Helfrich}W. Helfrich, Z. Naturforsch. C {\bf 28}, 693 (1973).

\bibitem{Canham} R. Canham, J. Theor. Biol.  {\bf 26}, 61 (1970).


\bibitem{PRC73:014309}P.-G. Reinhard, M. Bender, W. Nazarewicz, and T. Vertse, Phys. Rev. C {\bf 73}, 014309 (2006).

\bibitem{ZPhysA346:87}M. Durand, P. Schuck, X. Vifias,  Z. Phys. A {\bf 346}, 87 (1993).


\bibitem{curvedpasta} K. Nakazato, K. Iida, and K. Oyamatsu, Phys. Rev. C {\bf 83}, 065811 (2011).

\bibitem{Waffles}A. S. Schneider, D. K. Berry, C. M. Briggs, M. E. Caplan, and C. J. Horowitz, Phys. Rev. C {\bf 90}, 055805 (2014).

\bibitem{Puhka} M. Puhka {\it et al.}, Mol. Biol. Cell {\bf 23}, 2424 (2012).


\bibitem{nuclear_donuts} W. Nazarewicz, M. Bender, S. Cwiok, P.H. Heenen, A.T. Kruppa, P.-G. Reinhard, T. Vertse, Nuclear Physics A {\bf 701}, 165c (2002).

\bibitem{PhysRevC.69.045804} C. J. Horowitz, M. A. Perez-Garcia, and J.Piekarewicz, Phys. Rev. C {\bf 69}, 045804 (2004).

\bibitem{PhysRevC.70.065806} C. J. Horowitz, M. A. Perez-Garcia, J. Carriere, D. K. Berry, and J. Piekarewicz, Phys. Rev. C {\bf 70}, 065806 (2004).

\bibitem{PhysRevC.72.035801}C. J. Horowitz, M. A. Perez-Garcia, D. K. Berry, and J. Piekarewicz, Phys. Rev. C {\bf 72}, 035801 (2005).

\bibitem{PhysRevC.78.035806}C. J. Horowitz and D. K. Berry, Phys. Rev. C {\bf 78}, 035806 (2008).

\bibitem{PhysRevC.88.065807}A. S. Schneider, C. J. Horowitz, J. Hughto, and D. K. Berry, Phys. Rev. C {\bf 88}, 065807 (2013).

\bibitem{PhysRevC.86.055805}C. O. Dorso, P. A. Gimenez Molinelli, and J. A. Lopez, Phys. Rev. C {\bf 86}, 055805 (2012).

\bibitem{PhysRevC.89.055801}P. N. Alcain, P. A. Gimenez Molinelli, J. I. Nichols, and C. O. Dorso, Phys. Rev. C {\bf 89}, 055801 (2014).

\bibitem{Paraview} A. H. Squillacote, in The ParaView Guide: A Parallel Visualization Application (Kitware Inc., Clifton Park, New York, 2007), p. 366.

\bibitem{rmode}N. Andersson, K.D. Kokkotas, Int. J. Mod. Phys. D {\bf 10}, 381 (2001).

\bibitem{magnetar} A. L. Watts, T. E. Strohmayer, Advances in Space Research, {\bf 40}, 1446 (2007).

\bibitem{breakingstrain}C. J. Horowitz and Kai Kadau, Phys. Rev. Lett. {\bf 102}, 191102 (2009).

\end{thebibliography}
\end{document}